\documentclass[12pt]{article}
\usepackage{amsfonts}
\usepackage{lscape}
\usepackage[dvipdfmx]{graphicx}
\usepackage{color}

\newcommand{\Section}[1]%
{\section{#1}\setcounter{equation}{0}%
\setcounter{theorem}{0}}
\newtheorem{theorem}{Theorem}

\newtheorem{pro}[theorem]{Proposition}

\newtheorem{conjecture}[theorem]{Conjecture}

\def\re{\mathbb{R}}
\def\co{\mathbb{C}}

\def\na{\mathbb{N}}

{\par\noindent{\em #1:\ }}%
{~\rule{2mm}{2mm}\par\bigskip}

\begin{document}
\newpage\thispagestyle{empty}
{\topskip 2cm
\begin{center}
{\Huge\bf Quantization of Conductance\\\medskip in Quasi-Periodic Quantum Wires\\}
\bigskip\bigskip\bigskip
{\Large Tohru Koma\footnote{\label{fn}\small \it Department of Physics, Gakushuin University, Mejiro, 
Toshima-ku, Tokyo 171-8588, JAPAN,
{\small\tt e-mail: tohru.koma@gakushuin.ac.jp}},
Toru Morishita\footnote{\small \it Institute for Advanced Science, The University of Electro-communications,
1-5-1 Chofu-ga-oka, Chofu-shi, Tokyo 182-8585, JAPAN, 
{\small\tt e-mail: toru@pc.uec.ac.jp}} and Taro Shuya$^{\ref{fn},}$\footnote{\small \it Present address: 
Mitsubishi UFJ Trust Systems Co., Ltd. Konan, Minato-ku, Tokyo, 108-0075, JAPAN}
\\}
\end{center}
\vfil
\noindent
{\bf Abstract:} We study charge transport in the Peierls-Harper model 
with a quasi-periodic cosine potential. We compute the Landauer-type conductance of the wire. 
Our numerical results show the following: (i) When the Fermi energy lies in the absolutely 
continuous spectrum that is realized in the regime of the weak coupling,  
the conductance is quantized to the universal conductance. 
(ii) For the regime of localization that is realized for the strong coupling, the conductance is 
always vanishing irrespective of the value of the Fermi energy. 
Unfortunately, we cannot make a definite conclusion about the case with the critical coupling. 
We also compute the conductance of the Thue-Morse model. 
Although the potential of the model is not quasi-periodic, the energy spectrum is known to be a Cantor set 
with zero Lebesgue measure. Our numerical results for the Thue-Morse model also show the quantization of 
the conductance at many locations of the Fermi energy, except for the trivial localization regime. Besides, 
for the rest of the values of the Fermi energy, the conductance   
shows a similar behavior to that of the Peierls-Harper model with the critical coupling.      
\par
\noindent
\bigskip
\hrule
\bigskip
\vfil}

\newpage

\Section{Introduction}

Is the conductance of a quasi-periodic quantum wire quantized to the universal conductance 
when the Fermi level lies in the absolutely continuous spectrum?
Since a one-dimensional model with an ergodic potential is generally known to show a reflectionless property 
with probability one on the absolutely continuous spectrum for the scattering problem \cite{Kotani,Remling1,Remling2}, 
one can expect a positive answer to this question. 
However, there is a problem for determining whether or not the conductance is quantized to the universal value as follows: 
In theoretical approaches which rely on the Landauer formula \cite{Landauer,Landauer2}, 
the standard geometry of quantum wires consists of the one-dimensional sample and two ideal leads \cite{SJC,KLC,ROL1}. 
The ideal leads are connected to the two edges of the sample in order to measure the conductance of 
the sample. However, there appears a contact resistance at the edges of the sample. 
Thus, the pure conductance of the quasi-periodic wire cannot be detected in this method. 

In this paper, in order to avoid this difficulty, we propose an alternative approach as follows: 
Consider a sufficiently long quasi-periodic quantum wire with the periodic boundary condition. 
In order to induce a current in the wire, we adiabatically apply a constant voltage which is restricted to 
a subregion of the wire. We calculate the current through the left edge of the subregion by relying on 
a linear perturbation theory. As a result, the conductance given by the linear response coefficient 
is expressed in terms of a certain dynamical current-current correlation function. This is nothing but 
the Nakano-Kubo formula \cite{Nakano,Kubo,BVS,SB}. Thus, we do not use ideal leads. 

Even if starting from the Nakano-Kubo formula, it is a nontrivial problem 
whether or not the conductance is quantized to the universal value in quasiperiodic wires as follows. 
One might think that ballistic motions with the desired spectral and diffusion exponents \cite{SB2,Bellissard2} 
which are a consequence of the absolutely continuous spectrum always yield 
the quantization of the conductance. However, the value of the conductance is considerably sensitive to 
the local potential of the wire. Actually, the value of the conductance of the ideal wire with a single impurity 
strongly depends on the potential of the impurity. Besides, as we will see in the case of a quasi-periodic 
cosine potential, there appears a deviation from the quantized value of the conductance at the periodic boundary 
where the quasi-periodicity is slightly broken. As far as we know, there is no 
proof of the quantization of the conductance, and even no numerical demonstration. 
Thus, showing the quantization of the conductance in quasi-periodic wires is a nontrivial issue.

As a concrete example of the quasi-periodic quantum wires, we study the Peierls-Harper model \cite{Peierls,Harper}. 
The model first appeared in the work of Peierls \cite{Peierls} 
in which a tight-binding approximation was used for an electron system in both of a periodic potential and 
a magnetic field \cite{Harper}.
Our numerical results for this model show the following: (i) When the Fermi energy lies in the absolutely 
continuous spectrum that is realized in the regime of the weak coupling,  
the conductance is quantized to the universal conductance. 
(ii) For the regime of localization that is realized for the strong coupling, the conductance is 
always vanishing irrespective of the value of the Fermi energy. 
Unfortunately, we cannot make a definite conclusion about the case with the critical coupling. 

We also compute the conductance of the Thue-Morse tight-binding model \cite{Luck,Bellissard}. 
Although the potential of the model is not quasi-periodic, 
the energy spectrum is known to be a Cantor set with zero Lebesgue measure \cite{Luck,Bellissard,BBG,BG}. 
The same property appears in the Peierls-Harper model with the critical coupling. 
Our numerical results for the Thue-Morse model show the quantization of the conductance at many locations of the Fermi energy, 
except for the trivial localization regime. Besides, 
for the rest of the values of the Fermi energy, the conductance   
shows a similar behavior to that of the Peierls-Harper model with the critical coupling, 
although we cannot make a definite conclusion again.  

The present paper is organized as follows: In the next section, we give the precise definition of 
our model of the quantum wire, and derive the conductance formula by using the linear perturbation theory. 
Our numerical results are given in Sec.~\ref{Sec:NimericalCos} and Sec.~\ref{Sec:NimericalTM} 
for the Peierls-Harper and Thue-Morse models, respectively. 
Appendices~\ref{Appnd:LRF}-\ref{appendix:ProofconduRelationPeriodic} are devoted to 
the proofs of some statements and technical estimates. 

\Section{Model and Conductance Formula}

The outline of this section is as follows: 
We first describe our model whose Hamiltonian has a time-dependent potential term 
to induce a current in the quantum wire in Sec.~\ref{model}, 
and then recall the linear response theory to calculate the conductance in Sec.~\ref{LRT}.  
Finally, in Sec.\ref{C-Ccorrelation}, we express the conductance in terms of 
a certain dynamical current-current correlation function.  

\subsection{Lattice Electrons Driven by a Voltage Difference}
\label{model}

We consider a one-dimensional tight-binding model. The Hamiltonian $H_0$ acts on 
the wavefunction $\varphi(n)\in\co$, $n=1,2,\ldots, L$, as  
\begin{equation}
\label{H}
(H_0\varphi)(n)=-\varphi(n+1)-\varphi(n-1)+v(n)\varphi(n), 
\end{equation}
where $v(n)$ is a real-valued potential at the site $n$, and $L$ is the length of the chain. 
We impose the periodic boundary condition $\varphi(n+L)=\varphi(n)$ for the wavefunctions. 

In order to measure the current, we adiabatically apply a voltage difference $\mu$ 
on an interval $[j,j+\ell]\subset[1,L]$ with $j,\ell\in\na$. 
The corresponding characteristic function $\chi_{[j,j+\ell]}$ is given by  
$$
\chi_{[j,j+\ell]}(n):=\cases{1, & $n\in \{j,j+1,\ldots,j+\ell\}$; \cr 
                0, & otherwise. \cr}
$$
We also introduce a time-dependent function $W(t)$ as 
\begin{equation}
W(t):=-e^{\eta t}\chi_{[j,j+\ell]}\quad \mbox{for the time \ } t\in[-T,0],
\label{W(t)} 
\end{equation}
where $\eta>0$ is an adiabatic parameter, and $T$ is a positive number. 
We measure the current at the time $t=0$, after adiabatically applying the perturbation $\mu W(t)$ with the initial time $-T$.  
The total Hamiltonian $H(t)$ is given by 
$$
H(t):=H_0+\mu W(t). 
$$
 
\subsection{Linear Response Theory}
\label{LRT}

The time-dependent Schr\"odinegr equation is given by 
\begin{equation}
\label{tSchro}
i\frac{d}{dt}\Psi(t)=H(t)\Psi(t)
\end{equation}
for the wavefunction $\Psi(t)$. As usual, we introduce the time-evolution operator for 
the unperturbed Hamiltonian $H_0$ as 
$$
U_0(t,s):=\exp[-i(t-s)H_0]\quad \mbox{for \ } t,s\in\re.
$$
By using the linear response theory, one has the wavefunction $\Psi(0)$ at the time $t=0$ as 
\begin{equation}
\label{Psi(0)}
\Psi(0)=\Phi-i\mu \int_{-T}^0 ds\; U_0(0,s)W(s)U_0(s,0)\Phi + \mathcal{O}(\mu^2),
\end{equation}
where we have chosen the initial state $\Psi(-T)$ as \cite{Kato,Koma}
$$
\Psi(-T)=U_0(-T,0)\Phi
$$
with a normalized wavefunction $\Phi$. The derivation of (\ref{Psi(0)}) is given in Appendix~\ref{Appnd:LRF}. 

The current operator $J_j$ at the site $j$ is given by \cite{Mahan,Koma1}
\begin{equation}
\label{defJj}
(J_j\varphi)(n)=\cases{-i\varphi(j), & $n=j-1$; \cr 
i\varphi(j-1), & $n=j$;\cr 
                0, & otherwise, \cr}
\end{equation}
for a wavefunction $\varphi$. Therefore, the expectation value $I_j[\Phi]$ of the current for a given initial state $\Phi$ is 
\begin{eqnarray}
I_j[\Phi]&:=&\langle \Psi(0),J_j\Psi(0)\rangle\nonumber \\ 
&=&\langle \Phi,J_j\Phi\rangle-i\mu\int_{-T}^0 ds\; \langle \Phi,J_jU_0(0,s)W(s)U_0(s,0)\Phi\rangle\nonumber \\ 
&+&i\mu\int_{-T}^0 ds\; \langle U_0(0,s)W(s)U_0(s,0)\Phi,J_j\Phi\rangle+\mathcal{O}(\mu^2) 
\end{eqnarray}
for a small voltage difference $\mu$. We write 
$$
\chi_{[j,j+\ell]}(s):=e^{isH_0}\chi_{[j,j+\ell]}e^{-isH_0}. 
$$
As a result, the conductance which is derived as the linear response coefficient can be written as  
$$
g_j[\Phi]:=i\int_{-T}^0 ds\; e^{\eta s}\langle \Phi,[J_j,\chi_{[j,j+\ell]}(s)]\Phi\rangle,
$$
where we have used the expression (\ref{W(t)}) of $W(t)$ whose support is chosen so that the left edge site $j$ is the same 
as the site of the current operator $J_j$, i.e., we measure the current through the left edge of the support of the applied 
voltage difference. 

The total conductance for all the states below the Fermi level $E_{\rm F}$ is written as 
\begin{equation}
\label{gjetaTellN}
g_j(\eta,T,\ell,L):=\sum_{k:E_k\le E_{\rm F}}g_j[\Phi_k]=i\sum_{k:E_k\le E_{\rm F}}\int_{-T}^0 ds\; 
e^{\eta s}\langle \Phi_k,[J_j,\chi_{[j,j+\ell]}(s)]\Phi_k\rangle, 
\end{equation}
where $\Phi_k$ are the eigenstates of the unperturbed Hamiltonian $H_0$ 
with the energy eigenvalue $E_k$, i.e., they satisfy $H_0\Phi_k=E_k\Phi_k$. 
The conductance in the infinite-length limit $L\nearrow\infty$ is given by 
\begin{equation}
\label{gjinf}
g_j:=\lim_{\eta\searrow 0}\lim_{T\nearrow \infty}\lim_{\ell \nearrow \infty}
\lim_{L\nearrow \infty}g_j(\eta,T,\ell,L).
\end{equation}
We remark that the conductance formula (\ref{gjetaTellN}) is closely related to the Landauer-B\"uttiker formula 
in the work \cite{CorneanJensen} by Cornean and Jensen. 
Although their setting is totally different from ours, they justified 
the Landauer-B\"uttiker formula for a tight-binding model in a mathematically rigorous manner. 
See also a related work \cite{CGZ}. 

\subsection{Conductance as a Current-Current Correlation} 
\label{C-Ccorrelation}

As is well known, conductance (or conductivity) is often expressed in terms of a current-current correlation function. 
In this section, we show that the conductance (\ref{gjinf}) can be written also in terms of a current-current 
correlation function under a certain condition.   

We define 
\begin{equation}
\label{gjCCcor}
\hat{g}_j(\eta,L):=-i\sum_{k:E_k\le E_{\rm F}}\int_{-\infty}^0 ds\; 
s e^{\eta s}\langle \Phi_k,[J_j,J_j(s)]\Phi_k\rangle, 
\end{equation}
where 
$$
J_j(s):=e^{isH_0}J_je^{-isH_0}. 
$$
This is the well-known expression of the conductance in terms of the current-current correlation \cite{Nakano,Kubo}. 
We write 
\begin{equation}
\label{gjetaT}
g_j(\eta,T):=\lim_{\ell\nearrow \infty}\lim_{L\nearrow\infty}g_j(\eta,T,\ell,L).
\end{equation}

\begin{pro}
\label{pro:gjCCcor}
Suppose that the limit $g_j$ of (\ref{gjinf}) exists and that  
there exists a positive constant $\mathcal{C}_1$ which is independent of $j, \eta$ and $T$ such that 
the quantity $g_j(\eta,T)$ satisfies 
\begin{equation}
\label{assumgj}
|g_j(\eta,T)|\le \mathcal{C}_1
\end{equation}
for any $j,\eta$ and $T$. Then, the conductance $g_j$ of (\ref{gjinf}) is equal to 
the infinite-length limit of the conductance $\hat{g}_j(\eta,L)$ of (\ref{gjCCcor}) as   
\begin{equation}
\label{gj=JCorCor}
g_j=\lim_{\eta\searrow 0}\lim_{L\nearrow \infty}\hat{g}_j(\eta,L).
\end{equation} 
\end{pro}

The proof is given in Appendix~\ref{Appnd:PoofPropgjCCcor}. 

\bigskip

\noindent
{\it Remark:} (i) In Appendix~\ref{appendix:ProofconduRelationPeriodic},  
the relation (\ref{gj=JCorCor}) is justified in the case of periodic potentials without relying on 
the assumptions of Proposition~\ref{pro:gjCCcor}. 
\smallskip

\noindent
(ii) The physical meaning of the bound (\ref{assumgj}) is as follows: Even if starting from any initial 
time $-T$, the conductance observed at the time $t=0$ is bounded irrespective of the site $j$ and the cutoff $\eta$. 
Obviously, the value of the conductance fluctuates depending on the initial time $-T$. 
In order to eliminate such oscillations, we need to take the limit $T\nearrow\infty$ 
before taking the limit $\eta\searrow 0$.  
\medskip

\begin{theorem}
\label{theorem:periodic}
Suppose that the potential $v$ of the Hamiltonian $H_0$ of (\ref{H}) is a bounded periodic potential with 
period $p>0$, i.e., $v(n)$ satisfies $v(n+p)=v(n)$ for all sites $n$. 
Assume that all of the energy bands are separated by nonvanishing energy gaps, 
and that the Fermi level $E_{\rm F}$ lies inside a single band with a pair of two Fermi points 
with nonvanishing Fermi velocities. 
Then, the conductance $g_j$ in the infinite-length limit is quantized as 
\begin{equation}
g_j=\frac{1}{2\pi}. 
\end{equation}
\end{theorem}

In the calculations, if one includes the physical constants, electric charge $e$ and Planck constant $h$, 
then the result implies that the conductance is quantized to the universal value as $g_j=e^2/h$. 
The proof is given in Appendix~\ref{Appnd:PoofTheoremPeriodic}.  
\bigskip

\noindent
{\it Remark:} (i) Because of the translational invariance, the energy eigenvalues of the Hamiltonian $H_0$ are written 
as a function $E_{m,k}$ of the band index $m $ and the wavenumber $k$. Then, the Fermi velocities $v_{\rm F}$ are given by 
$$ 
v_{\rm F}=\frac{d}{dk}E_{m,k}\Bigr|_{E_{m,k}=E_{\rm F}}
$$
at each Fermi point.
\smallskip

\noindent
(ii) We can relax the assumption on the energy bands of the Hamiltonian $H_0$. Roughly speaking, 
the existence of the absolutely continuous spectrum near the Fermi energy with nonvanishing Fermi velocities 
yields the quantization of the conductance.  
\medskip

Next consider the Peierls-Harper model \cite{Peierls,Harper}, whose potential is a quasi-periodic cosine potential  
in Conjecture~\ref{conjecture} below. Mathematicians often call the Hamiltonian almost Mathieu operator.  
For a review of the results about the model or the operator, see \cite{MJ}. For a multifractal analysis of 
the wavefunctions, see, for example, \cite{TK,KST,JK}.  

\begin{conjecture}
\label{conjecture}
Suppose that the potential $v$ of the Hamiltonian $H_0$ of (\ref{H}) is 
the quasi-periodic potential $v(n)=U\cos(2\pi \omega n)$ with $-2<U<2$ and 
an irrational real $\omega$, and assume that the Fermi level lies ``inside" the absolutely continuous spectrum of 
the Hamiltonian $H_0$ in the infinite-length limit. 
Then, the conductance $g_j$ in the infinite-length limit is quantized as 
\begin{equation}
g_j=\frac{1}{2\pi}. 
\end{equation}
\end{conjecture} 
\medskip

\noindent
{\it Remark:} Although we cannot define ``inside" in a mathematical rigorous manner, we can expect 
the following: If the spectrum of the quasi-periodic system in Conjecture~\ref{conjecture} can be obtained from  
the spectrum of a periodic system in the infinite limit of the period \cite{JM}, then the conductance of the quasi-periodic 
system exhibits the same quantization as that of the periodic system. 
Our numerical results in Sec.~\ref{Sec:NimericalCos} below strongly support this conjecture.

\Section{Numerical Results}
\label{Sec:NumericalR}

Before proceeding to the numerical computations of the conductance for the present model with concrete potentials, 
we prepare a useful expression of the conductance for the computations. 

{From} the expression (\ref{defJj}) of the current operator $J_j$, one can easily obtain 
the following expression of the conductance $\hat{g}_j(\eta,L)$ of (\ref{gjCCcor}): 
\begin{equation}
\hat{g}_j(\eta,L)=\sum_{k:E_k\le E_{\rm F}}\sum_{k':E_{k'}>E_{\rm F}}
|\Phi_k(j-1)\Phi_{k'}(j)-\Phi_k(j)\Phi_{k'}(j-1)|^2\frac{4\eta(E_{k'}-E_k)}{[(E_{k'}-E_k)^2+\eta^2]^2},
\end{equation}
where $\Phi_k(j)$ is the amplitude of the normalized eigenstate $\Phi_k$ at the site $j$, and we have used 
the fact that all of the eigenstates $\Phi_k$ of the unperturbed Hamiltonian $H_0$ are taken to be a real-valued 
function because the Hamiltonian $H_0$ is a real symmetric matrix. 

We remark that a similar formula 
which is based on the $C^\ast$ algebra framework \cite{BVS,SB,Bellissard2} 
was applied to a one-dimensional incommensurate bilayer system \cite{CCL}. 
Although the authors of Ref.~\cite{CCL} numerically computed 
the conductivity, their aim was different from ours. More precisely, they did not address the issue of quantization 
of the conductance when the Fermi level lies in the absolutely continuous spectrum. 
See also related articles, \cite{Prodan} about a computational method for disordered systems, and 
\cite{ProdanBellissard} about a numerical computation for a current-current correlation.

\subsection{Quasi-Periodic Cosine Potential}
\label{Sec:NimericalCos}

Consider first the Peierls-Harper model as a concrete example. 
The potential term in the Hamiltonian $H_0$ of (\ref{H}) is the quasi-periodic cosine potential, 
\begin{equation}
v(n)=U\cos(2\pi \omega n),
\end{equation}
where $U\in\re$ is the strength of the potential. We choose $\omega=(\sqrt{5}+1)/{2}$. 
Since this $\omega$ is irrational, the potential $v$ is quasi-periodic.   
As is well known, when varying the strength $U$ of the cosine potential, 
the spectrum of the Hamiltonian $H_0$ exhibits the following three phases \cite{AA,Kohmoto,MJ}: 
\begin{enumerate} 
\item For $|U|<2$, the spectrum of the Hamiltonian $H_0$ is purely absolutely continuous. 
\item For $|U|>2$, the spectrum is pure point, i.e., the whole spectrum shows localization of the wavefunctions. 
\item For the critical value $U=\pm 2$, the spectrum is purely singular continuous.  
\end{enumerate}

\begin{figure}
\centering
\includegraphics[width=11.5cm]{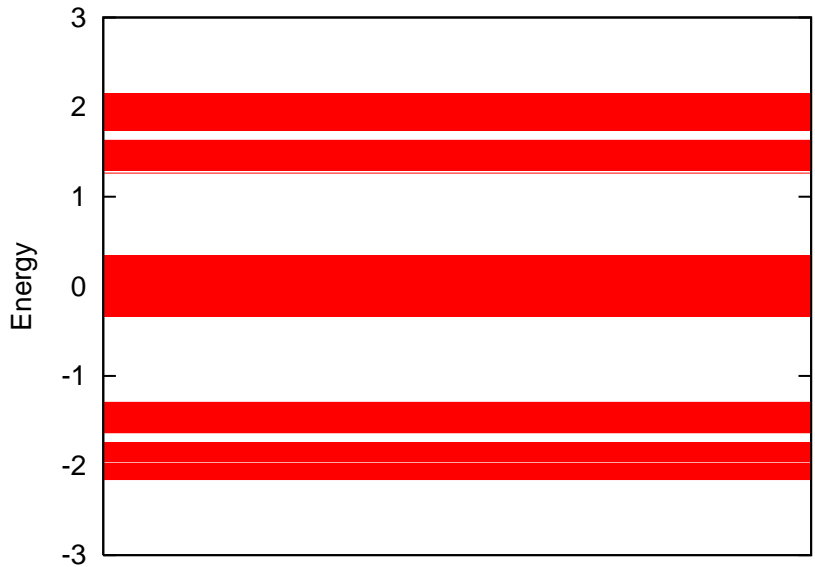}
\caption{The spectrum of the Peierls-Harper model with the strength $U=1$ of the cosine potential.}
\label{spectrumU1}
\end{figure}

\begin{figure}
\centering
\includegraphics[width=11.5cm]{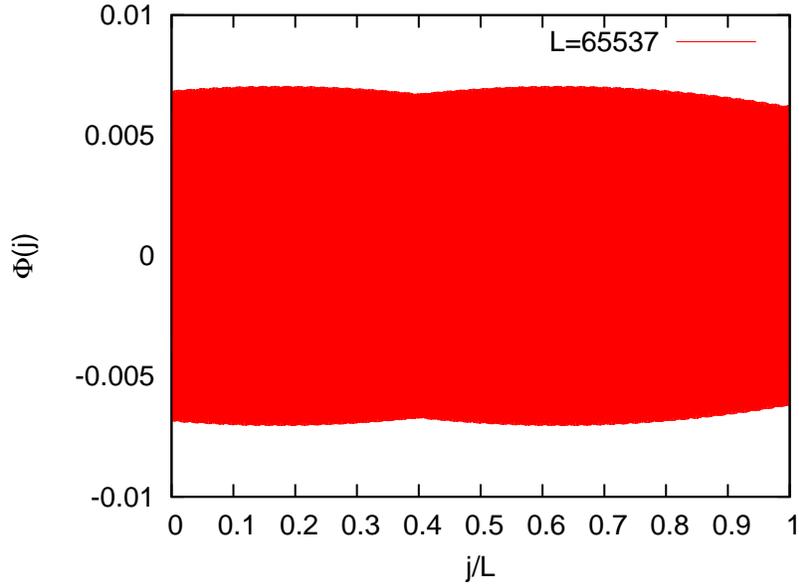}
\caption{The site-dependence of the real-valued amplitude of a typical energy eigenfunction $\Phi(j)$ 
with the energy eigenvalue $E=-2.6717\times 10^{-5}$ near the Fermi energy $E_{\rm F}=0$ in the Peierls-Harper model 
with the strength $U=1$ of the cosine potential.}
\label{wavefunctionU1}
\end{figure}

\begin{figure}
\centering
\includegraphics[width=11.5cm]{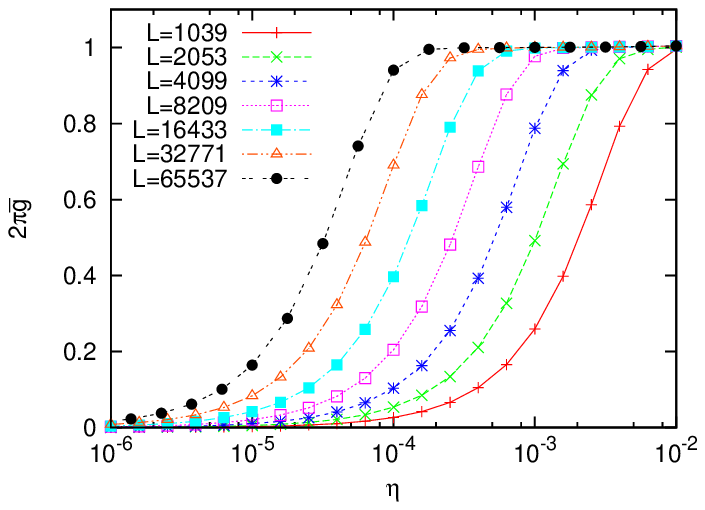}
\caption{The cutoff-dependence of the averaged conductance $\overline{g}$ 
for the Peierls-Harper model with the strength $U=1$ of the cosine potential. 
The Fermi energy are chosen to be $E_{\rm F}=0$.}
\label{cuoffU1}
\end{figure}

\begin{figure}
\centering
\includegraphics[width=11.5cm]{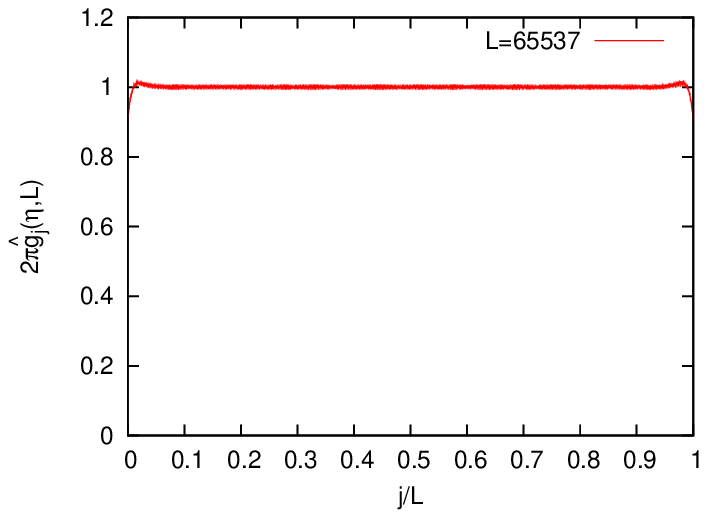}
\caption{The site-dependence of the conductance $\hat{g}_j(\eta,L)$ 
for the Peierls-Harper model with the strength $U=1$ of the cosine potential.  
The Fermi energy and the cutoff are, respectively, chosen to be $E_{\rm F}=0$ and $\eta=1.0\times 10^{-3}$.}
\label{siteU1}
\end{figure}

To begin with, we define the averaged conductance $\overline{g}$ and the standard deviation $\delta g$ by 
$$
\overline{g}:=\frac{1}{L}\sum_{j=1}^L \hat{g}_j(\eta,L)
\quad \mbox{and} \quad 
\delta g:=\sqrt{\frac{1}{L}\sum_{j=1}^L[\hat{g}_j(\eta,L)-\overline{g}]^2}.
$$
For $U=1$, the band spectrum is shown in Fig.~\ref{spectrumU1}, and  
one can see that a typical wavefunction is extended as shown in Fig.~\ref{wavefunctionU1}.  
Figure~\ref{cuoffU1} shows the cutoff-dependence of the averaged conductance $\overline{g}$ in the case with $U=1$. 
For a fixed small value of the cutoff $\eta$, the averaged conductance $\overline{g}$ converges to unity as the length of the wire increases.  
Figure~\ref{siteU1} shows the site-dependence of the conductance in the case with $U=1$. 
The conductance is quantized to unity with the mean value $2\pi\overline{g}=1.00040$ 
and the standard deviation $2\pi\delta g=6.70\times 10^{-3}$, 
where we impose the periodic boundary condition. 
As seen in Fig.~\ref{siteU1}, the deviations from the quantized value of the conductance at the boundary 
are larger than those at the bulk region. 
Clearly, the quasi-periodicity is broken at the boundary due to the periodic boundary condition, 
although we choose the length $L$ of the wire to be a prime number. 
Thus, the deviations from the quantized value of the conductance at the boundary can be interpreted as 
the effect of the periodic boundary condition. In other words, the contact resistance appears at the boundary 
as mentioned in Introduction. Thus, our numerical results strongly support 
the validity of Conjecture~\ref{conjecture}. 
 

\begin{figure}
\centering
\includegraphics[width=11.5cm]{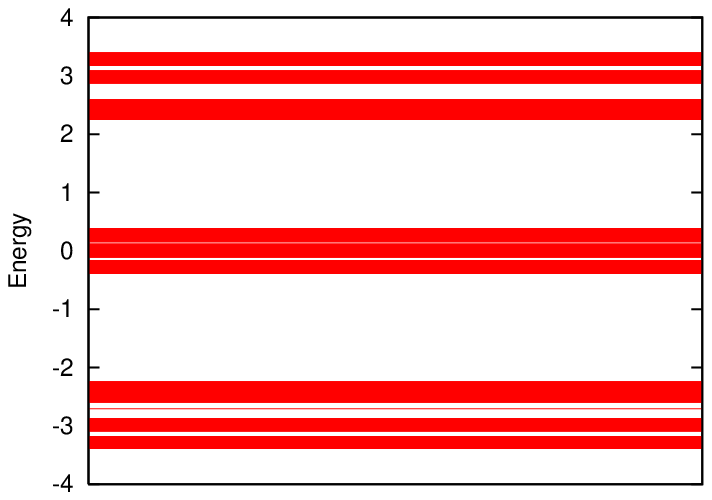}
\caption{The spectrum of the Peierls-Harper model with the strength $U=3$ of the cosine potential.}
\label{spectrumU3}
\end{figure}

\begin{figure}
\centering
\includegraphics[width=11.5cm]{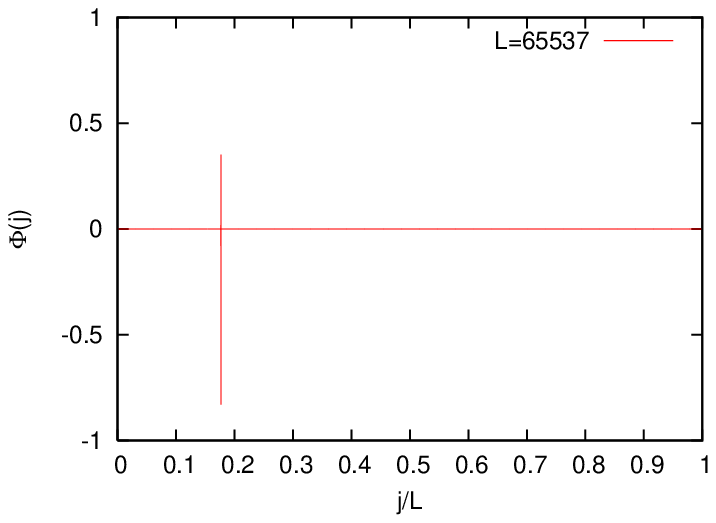}
\caption{A typical wavefunction $\Phi(j)$ with the energy eigenvalue $E=2.222697\times 10^{-5}$
near the Fermi energy $E_{\rm F}=0$ in the Peierls-Harper model with the strength $U=3$ of the cosine potential.}
\label{wavefunctionU3}
\end{figure}

\begin{figure}
\centering
\includegraphics[width=11.5cm]{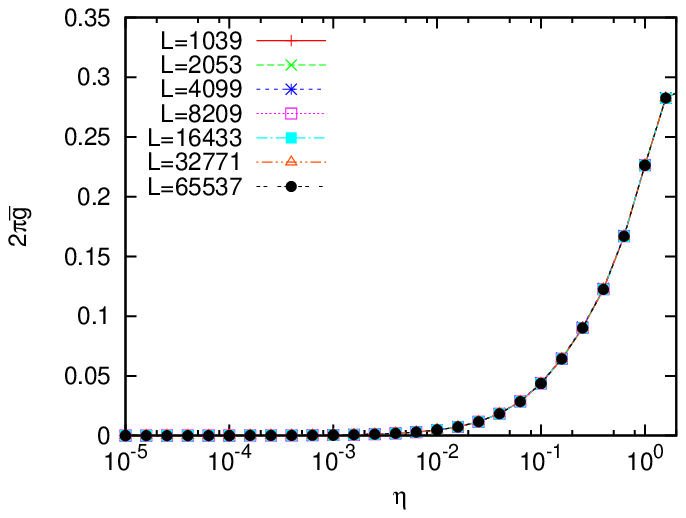}
\caption{The cutoff-dependence of the averaged conductance $\overline{g}$ for the Peierls-Harper model 
with the strength $U=3$ of the cosine potential. The Fermi energy is chosen to be $E_{\rm F}=0$.}
\label{cutoffU3}
\end{figure}

\begin{figure}
\centering
\includegraphics[width=11.5cm]{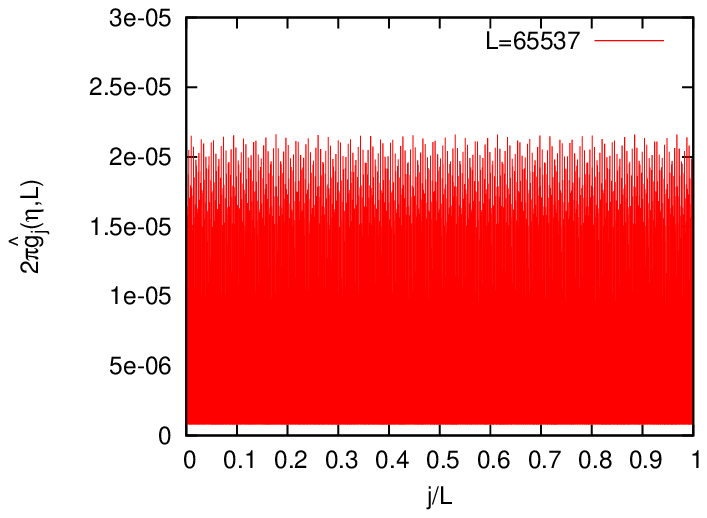}
\caption{The site-dependence of the conductance $\hat{g}_j(\eta,L)$ 
for the Peierls-Harper model with the strength $U=3$ of the cosine potential. 
The Fermi energy and the cutoff are, respectively, chosen to be $E_{\rm F}=0$ and $\eta=1\times 10^{-5}$.}
\label{siteU3}
\end{figure}

For $U=3$, the spectrum of the Hamiltonian is shown in Fig.~\ref{spectrumU3}. 
One can see that a typical wavefunction is localized as in Fig.~\ref{wavefunctionU3}. 
Clearly, one can expect that the conductance in this case is vanishing 
because all of the wavefunctions are localized. Actually, 
for a fixed small value of the cutoff $\eta$, the averaged conductance $\overline{g}$ converges to zero 
as the length of the wire increases as seen in Fig.~\ref{cutoffU3}. 
Besides, the conductance $\hat{g}_j(\eta,L)$ is almost vanishing at all of the sites 
for large lengths of the wire as in Fig.~\ref{siteU3}.  

\begin{figure}
\centering
\includegraphics[width=11.5cm]{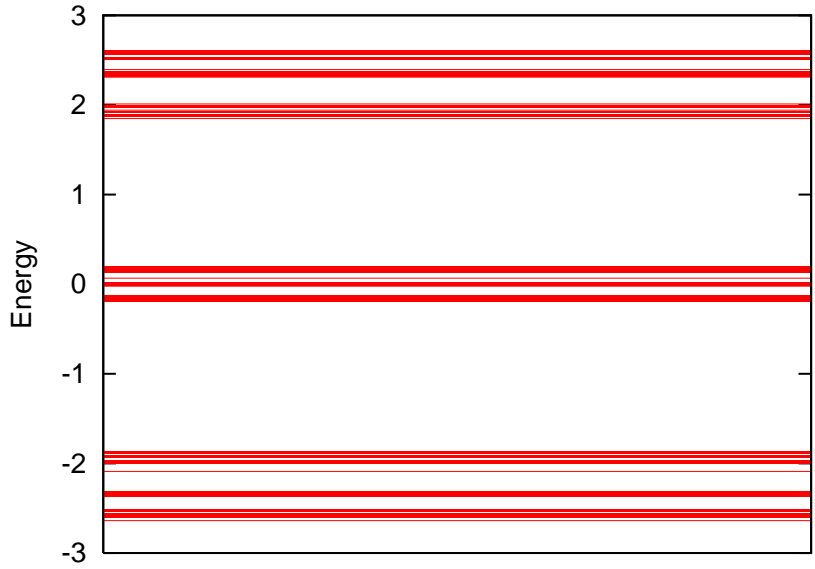}
\caption{The spectrum of the Peierls-Harper model with the strength $U=2$ of the cosine potential.}
\label{spectrumU2}
\end{figure}

\begin{figure}
\centering
\includegraphics[width=11.5cm]{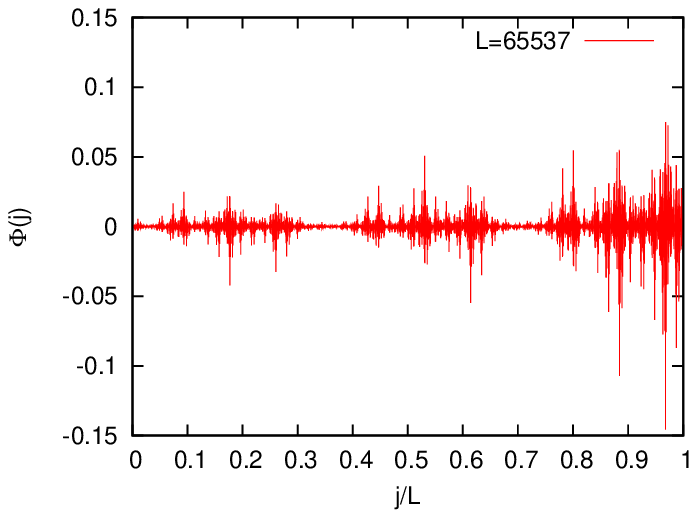}
\caption{A typical wavefunction $\Phi(j)$ with the energy eigenvalue is $E=-1.4593479\times 10^{-9}$
near the Fermi energy $E_{\rm F}=0$ in the Peierls-Harper model with the strength $U=2$ of the cosine potential.}
\label{wavefunctionU2}
\end{figure}

\begin{figure}
\centering
\includegraphics[width=11.5cm]{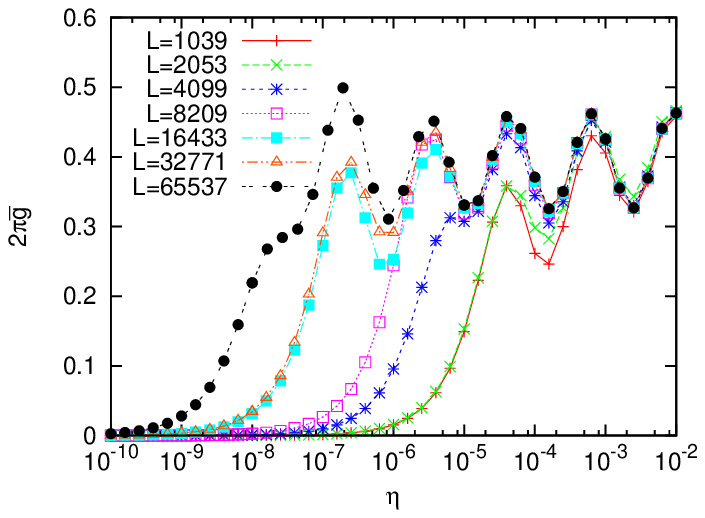}
\caption{The cutoff-dependence of the averaged conductance $\overline{g}$ 
for the Peierls-Harper model with the strength $U=2$ of the cosine potential.  
The Fermi energy is chosen to be $E_{\rm F}=0$.}
\label{cutoffU2}
\end{figure}

\begin{figure}
\centering
\includegraphics[width=11.5cm]{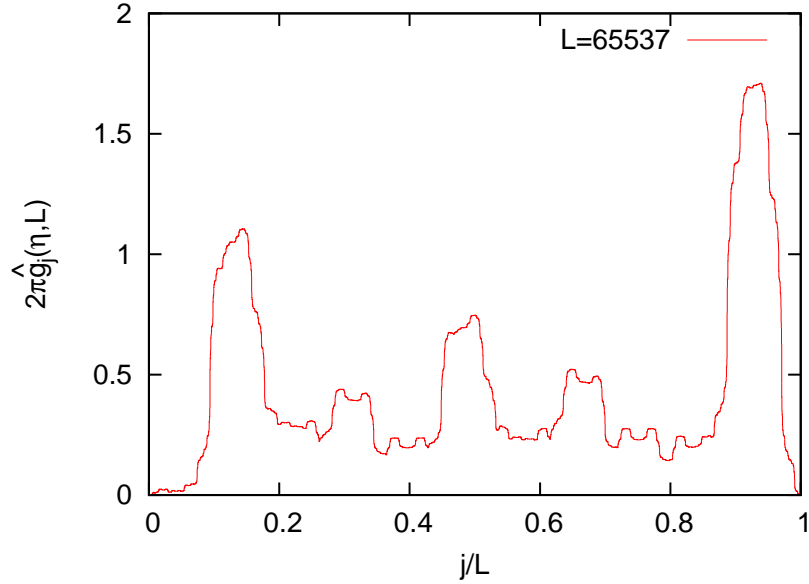}
\caption{The site-dependence of the conductance $\hat{g}_j(\eta,L)$ 
for the Peierls-Harper model with the strength $U=2$ of the cosine potential. 
The Fermi energy and the cutoff are chosen to be $E_{\rm F}=0$ and $\eta=1.17877\times 10^{-7}$, respectively.}
\label{siteU2}
\end{figure}

The rest is the case with the critical value $U=\pm 2$ of the strength of the cosine potential.  
In the following, we will consider the case $U=2$ only. 
The spectrum of the Hamiltonian $H_0$ is shown in Fig.~\ref{spectrumU2}. 
The singular continuous character of the spectrum emerges as a sparse structure. 
The peculiarity also emerges as the fractal character of the wavefunctions as seen in Fig.~\ref{wavefunctionU2}. 
Figure~\ref{cutoffU2} shows the cutoff-dependence of the averaged conductance $\overline{g}$ in the case with the critical value $U=2$. 
Clearly, in contrast to the above two cases, the conductance in this case does not converge to a value 
for a fixed small value of the cutoff $\eta$ as the length of the wire increases. 
Figure~\ref{siteU2} shows the site-dependence of the conductance $\hat{g}_j(\eta,L)$. 
The value of the cutoff $\eta$ is 
chosen to be $\eta=1.17877\times 10^{-7}$ which gives the maximum value of the averaged conductance $\overline{g}$ 
in Fig.~\ref{cutoffU2} with the length $L=65537$ of the wire.  
The value of the conductance $\hat{g}_j(\eta,L)$ strongly depends on the site $j$ of the wire, 
and fluctuates throughout the whole region. 
Since the Lebesgue measure of the singular continuous spectrum is vanishing, 
one can expect that the contribution of the density of the states to the electrical current is also 
vanishing. Therefore, the vanishing of the conductance can be expected. But our numerical results 
are considerably subtle for determining whether or not the conductance is quantized to a value. 

We also remark that in this critical case, the conductance is expected to show 
a power-law decay \cite{Kohmoto2,SutherlandKohmoto,IRT,RTM}   
with the length of the chain. Unfortunately, we have not been able to determine the length-dependence of the conductance 
by using our approach. Namely, our approach has not been able to give a conclusive answer to this issue.
In passing, as to anomalous diffusion exponents in a more general setting, see, e.g., \cite{KKL,DGLQ}.

\newpage

\subsection{Thue-Morse Potential}
\label{Sec:NimericalTM}

Next we treat the Thue-Morse Potential \cite{Luck,Bellissard,ROL1,ROL2,CJ}. 
In order to define the Thue-Morse potential, let us consider the Thue-Morse substitution $\varsigma$, 
which is defined by 
$\varsigma(A)=AB$ and $\varsigma(B)=BA$, where the symbols $A$ and $B$ denote $A$ and $B$ atoms, respectively. 
Repeated substitutions give the following sequence: 
$$
A\rightarrow AB \rightarrow ABBA \rightarrow ABBABAAB \rightarrow 
ABBABAABBAABABBA \rightarrow \cdots
$$
For the $(N+1)$-th generation with $N\ge 1$, we write 
$$
S=S_1S_2S_3\cdots S_{L-1}S_L,
$$
where the length $L$ of the chain is given by $L=2^N$. Then, the Thue-Morse potential is defined by 
\begin{equation}
v(n):=\cases{+1, & if $S_n=A$; \cr 
                -1, & if $S_n=B$. \cr}
\end{equation}
Here, we have chosen $U=1$ as the strength of the coupling. 

\begin{figure}
\centering
\includegraphics[width=11.5cm]{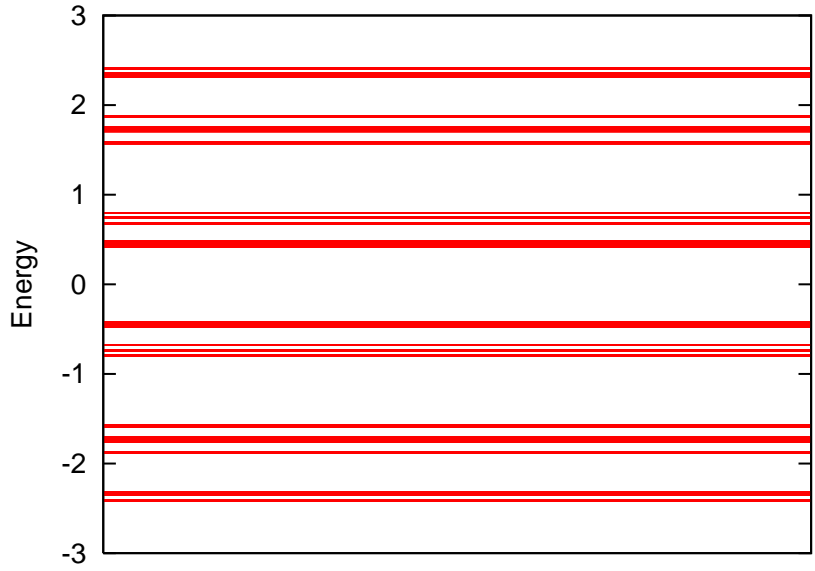}
\caption{The spectrum of the Thue-Morse model.}
\label{spectrumTM}
\end{figure}

\begin{figure}
\centering
\includegraphics[width=11.5cm]{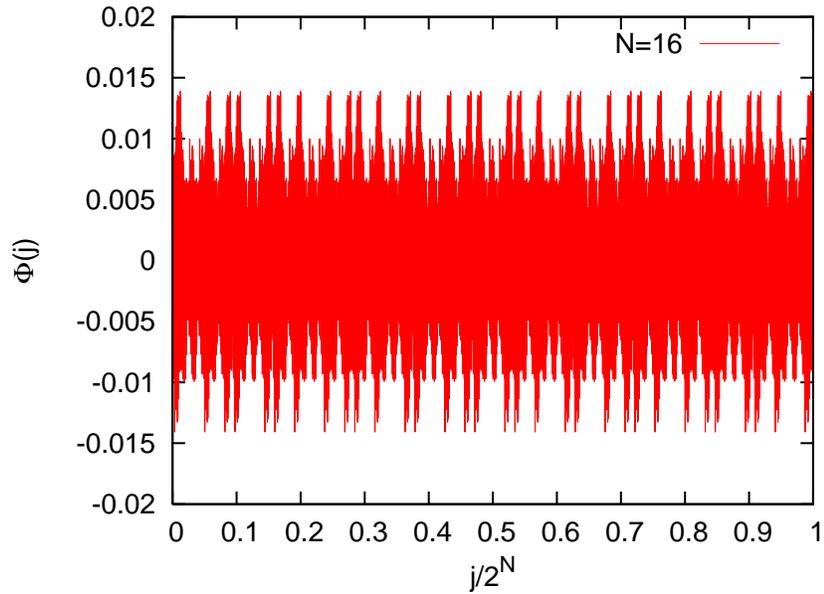}
\caption{A typical wavefunction $\Phi(j)$ with the energy eigenvalue $E=0.42714456$ 
near the Fermi energy $E_{\rm F}=0.42715$ in the Thue-Morse model.}
\label{wavefunctionextendTM}
\end{figure}

\begin{figure}
\centering
\rotatebox{0}{
\includegraphics[width=11.5cm]{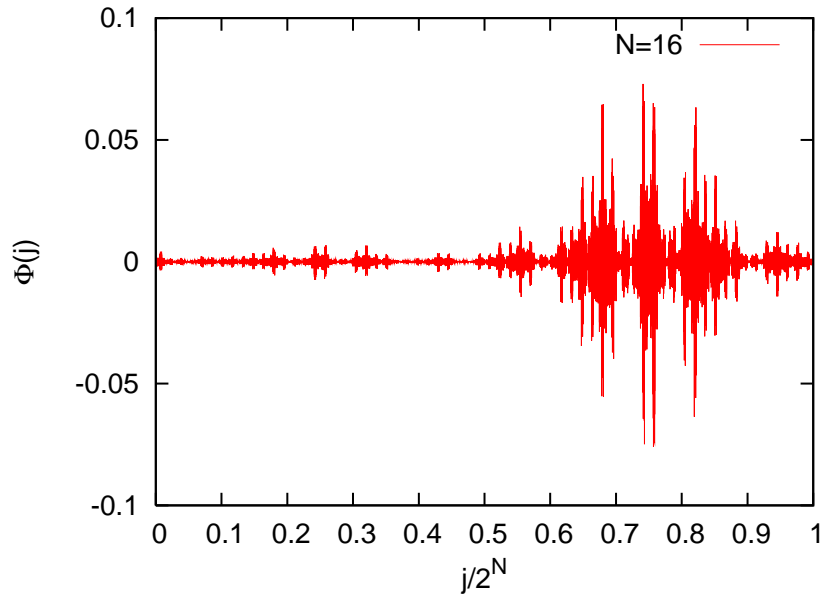}}
\caption{A typical wavefunction $\Phi(j)$ with the energy eigenvalue $E=0.73607815$ 
near the Fermi energy $E_{\rm F}=0.736078$ in the Thue-Morse model.}
\label{wavefunctioncriticalTM}
\end{figure}

Figure~\ref{spectrumTM} show the spectrum of the Thue-Morse model. 
This resembles the spectrum of the Peierls-Harper model with the critical coupling $U=2$ 
in the sparse structure as seen in Fig.~\ref{spectrumU2}. 
This is because the spectrum is a Cantor set with zero Lebesgue measure as mentioned in Introduction. 
However, depending on the locations of 
the Fermi energy $E_{\rm F}$, there often appear the wavefunctions which 
exhibit a different character from the critical wavefunctions in the Peierls-Harper model 
with the critical coupling as seen in Fig.~\ref{wavefunctionextendTM}. 
The wavefunction looks more like extended than critical. These extended states were already 
observed in numerical computations \cite{ROL1,ROL2,CJ}, and later their existence was confirmed by \cite{LQY} 
in a mathematically rigorous manner. 
The wavefunctions which look like a critical 
wavefunction also appear depending on the values of the Fermi energy $E_{\rm F}$. 
Figure~\ref{wavefunctioncriticalTM} shows such a typical wavefunction  
that closely resembles to the critical wavefunction in Fig.~\ref{wavefunctionU2}. 

\begin{figure}
\centering
\includegraphics[width=11.5cm]{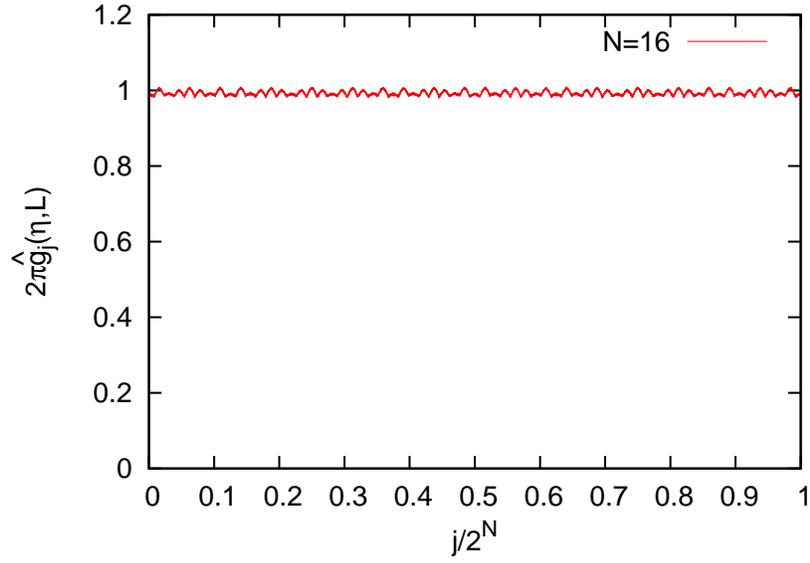}
\caption{The site-dependence of the conductance $\hat{g}_j(\eta,L)$ for the Thue-Morse model. 
The Fermi energy and the cutoff are chosen to be $E_{\rm F}=0.42715$ and $\eta=2.51188\times 10^{-5}$, respectively.}
\label{conductanceextendTM}
\end{figure}

\begin{figure}
\centering
\includegraphics[width=11.5cm]{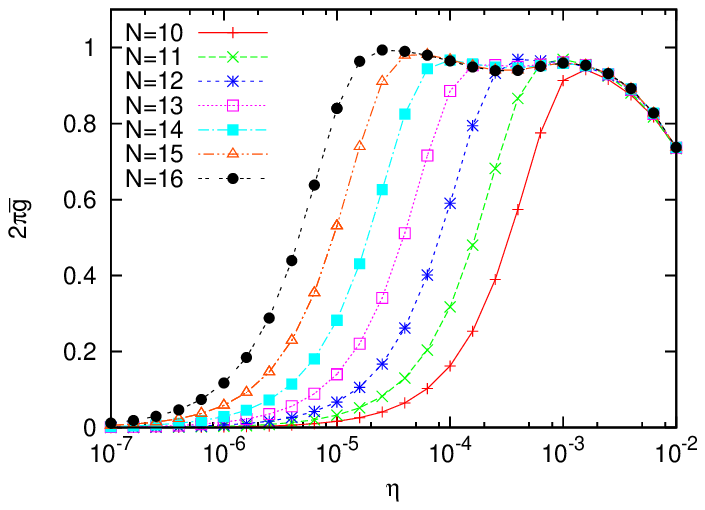}
\caption{The cutoff-dependence of the averaged conductance $\overline{g}$ for the Thue-Morse model. 
The Fermi energy is chosen to be $E_{\rm F}=0.42715$.}
\label{cutoffextendTM}
\end{figure}

In the case of the extended-like wavefunctions with the Fermi energy $E_{\rm F}=0.42715$, 
the conductance $\hat{g}_j(\eta,L)$ is quantized to the universal value 
with the mean value $2\pi \overline{g}= 0.99339$ and the standard deviation $2\pi \delta g= 6.29\times 10^{-3}$ 
as seen in Fig.~\ref{conductanceextendTM}. The value of the cutoff $\eta$ is chosen to be $\eta=2.5119\times 10^{-5}$
which gives the maximum value of the averaged conductance $\overline{g}$ in Fig.~\ref{cutoffextendTM} 
with the length $L=2^N$ of the wire with $N=16$. In contrast to the case of the Peierls-Harper model 
with the strength $U=1$ of the cosine potential, it does not look like there appears a contact resistance at the boundary, 
which is due to the periodic boundary condition. 

\begin{figure}
\centering
\includegraphics[width=11.5cm]{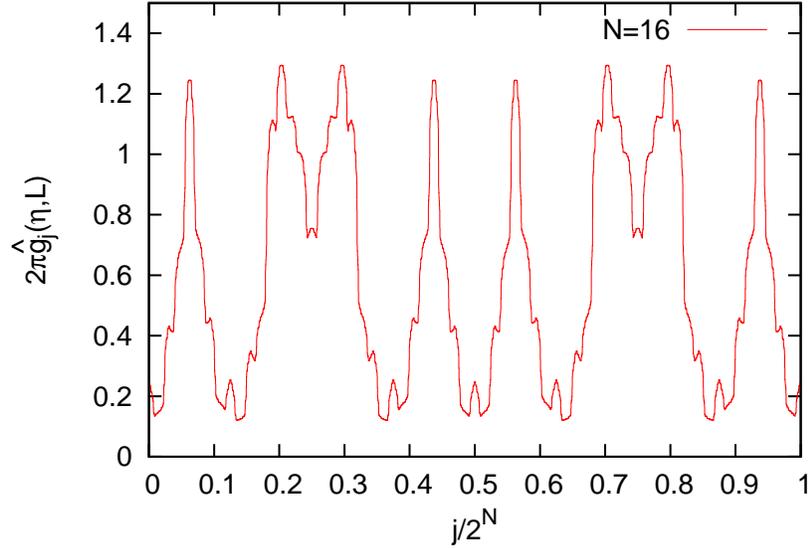}
\caption{The site-dependence of the conductance $\hat{g}_j(\eta,L)$ for the Thue-Morse model. 
The Fermi energy and the cutoff are chosen to be $E_{\rm F}=0.736078$ and $\eta=3.981\times 10^{-7}$, respectively.}
\label{conductancecriticalTM}
\end{figure}

\begin{figure}
\centering
\includegraphics[width=11.5cm]{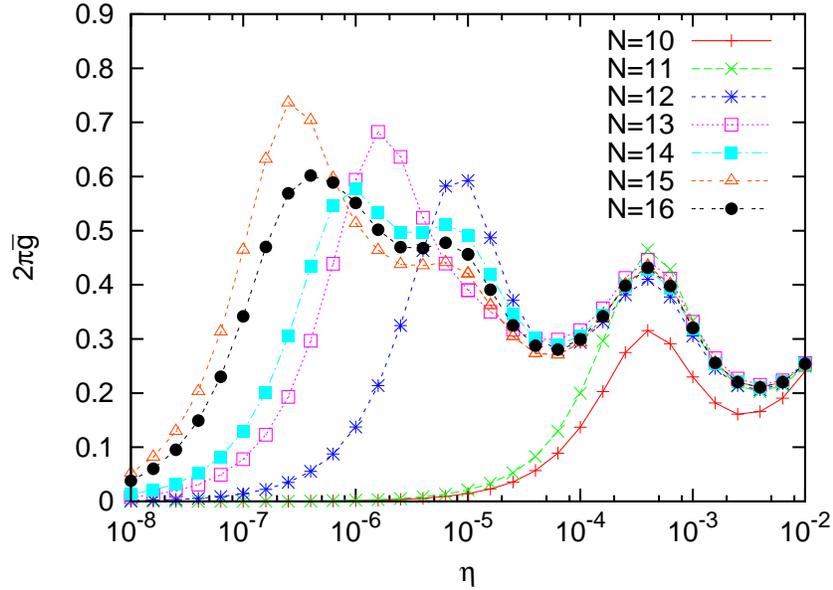}
\caption{The cutoff-dependence of the averaged conductance $\overline{g}$ for the Thue-Morse model. 
The Fermi energy is chosen to be $E_{\rm F}=0.736078$.}
\label{cutoffcriticalTM}
\end{figure}

The case of the critical-like wavefunctions with the Fermi energy $E_{\rm F}=0.736078$ 
does not show the quantization of the conductance as shown in Fig.~\ref{conductancecriticalTM}. 
The value of the cutoff $\eta$ is chosen to be $\eta=3.981\times 10^{-7}$ which gives 
the maximum value of the averaged conductance $\overline{g}$ in Fig.~\ref{cutoffcriticalTM}
with the length $L=2^N$ of the wire with $N=16$. 
Clearly, one can see that the conductance fluctuates depending on the site $j$.

The observations in Figs.~\ref{conductanceextendTM} and \ref{conductancecriticalTM} are consistent with the cutoff-dependence of the conductance 
in Figs.~\ref{cutoffextendTM} and \ref{cutoffcriticalTM}. In the former case, the conductance 
converges to unity for a fixed small value of the cutoff $\eta$.    
The latter case shows a similar behavior to that of the Peierls-Harper model with the critical 
coupling in Fig.~\ref{cutoffU2}. Although the Lebesgue measure of the spectrum is vanishing, 
the conductance is quantized to the universal value for the extended-like wavefunctions.  
The reason can be interpreted that the Hausdorff dimension of the spectrum of the Thue-Morse model \cite{LQ,LQY} 
is much larger than that of the Peierls-Harper model with the critical coupling. 
In other words, the density of the spectrum of the extended-like states in the Thue-Morse model 
is large enough compared to that of the critical states in the Peierls-Harper model with the critical coupling, 
although both of the two models exhibit the spectrum of a Cantor set with zero Lebesgue measure. 
In consequence, the extended-like states in the Thue-Morse model yield the quantization of the conductance.  

\newpage

\appendix

\Section{Derivation of of Linear Response Formula (\ref{Psi(0)})}
\label{Appnd:LRF}

In this appendix, we recall the derivation of the expansion  (\ref{Psi(0)}) for wavefunctions 
by following Kato \cite{Kato}. 

Let $\Psi(s)$ be a solution of the Schr\"odinger equation (\ref{tSchro}). Then, one has 
\begin{eqnarray*}
\frac{d}{ds}U_0(t,s)\Psi(s)&=&\left[\frac{d}{ds}U_0(t,s)\right]\Psi(s)+U_0(t,s)\frac{d}{ds}\Psi(s)\\
&=&iU_0(t,s)H_0\Psi(s)-iU_0(t,s)H(s)\Psi(s)\\
&=&-i\mu U_0(t,s)W(s)\Psi(s).
\end{eqnarray*}
Integrating over $s$ from $-T$ to $t$, one obtains 
$$
\Psi(t)-U_0(t,-T)\Psi(-T)=-i\mu\int_{-T}^t ds\; U_0(t,s)W(s)\Psi(s).
$$
Therefore, we have 
$$
\Psi(t)=U_0(t,-T)\Psi(-T)+\mathcal{O}(\mu)=U_0(t,0)\Phi+\mathcal{O}(\mu),
$$
where we have used the initial condition, $\Psi(-T)=U_0(-T,0)\Phi$. 
Substituting this into the above integrand, we obtain the desired result (\ref{Psi(0)}).

\Section{Proof of Proposition~\ref{pro:gjCCcor}}
\label{Appnd:PoofPropgjCCcor}

In this appendix, we prove that the conductance $g_j$ of (\ref{gjinf}) is equal to the standard form, 
which is expressed in terms of the current-current correlation function as in (\ref{gj=JCorCor}) with 
(\ref{gjCCcor}), under the assumption (\ref{assumgj}).   

Consider the summand in the right-hand side of (\ref{gjetaTellN}).  
By integration by parts, one has 
\begin{eqnarray}
& &\int_{-T}^0 ds\; 
e^{\eta s}\langle \Phi_k,[J_j,\chi_{[j,j+\ell]}(s)]\Phi_k\rangle \nonumber \\
&=&\left[s e^{\eta s}\langle \Phi_k,[J_j,\chi_{[j,j+\ell]}(s)]\Phi_k\rangle\right]_{-T}^0
-\int_{-T}^0 ds\; s\frac{d}{ds}
e^{\eta s}\langle \Phi_k,[J_j,\chi_{[j,j+\ell]}(s)]\Phi_k\rangle \nonumber \\
&=&T e^{-\eta T}\langle \Phi_k,[J_j,\chi_{[j,j+\ell]}(-T)]\Phi_k\rangle
-\int_{-T}^0 ds\; \eta s
e^{\eta s}\langle \Phi_k,[J_j,\chi_{[j,j+\ell]}(s)]\Phi_k\rangle \nonumber \\
& &-\int_{-T}^0 ds\; s
e^{\eta s}\langle \Phi_k,[J_j,\frac{d}{ds}\chi_{[j,j+\ell]}(s)]\Phi_k\rangle.
\label{CorJchi} 
\end{eqnarray}
In the following three subsections, we will estimate the three terms in the right-hand side. 

\subsection{Estimating the first term in the right-hand side of (\ref{CorJchi})}
\label{firsttermCorJchi}

To begin with, we note that 
\begin{equation}
\label{derivchis}
\frac{d}{ds}\chi_{[j,j+\ell]}(s)=e^{isH_0}(J_j-J_{j+\ell+1})e^{-isH_0}=J_j(s)-J_{j+\ell+1}(s)
\end{equation}
which is derived from the definition of the current operator $J_j$. Integrating over $s$, one has 
\begin{equation}
\label{chijjell1s}
\chi_{[j,j+\ell]}(s)=\chi_{[j,j+\ell]}(0)+\int_0^s d\tau\; [J_j(\tau)-J_{j+\ell+1}(\tau)].
\end{equation}
Substituting this into the first term in the right-hand side of (\ref{CorJchi}), we have 
\begin{eqnarray}
& &T e^{-\eta T}\langle \Phi_k,[J_j,\chi_{[j,j+\ell]}(-T)]\Phi_k\rangle \nonumber \\
&=&T e^{-\eta T}\langle \Phi_k,[J_j,\chi_{[j,j+\ell]}]\Phi_k\rangle
+T e^{-\eta T}\int_0^{-T}d\tau\; \langle \Phi_k,[J_j,J_{j}(\tau)]\Phi_k\rangle \nonumber \\
& &-T e^{-\eta T}\int_0^{-T}d\tau\; \langle \Phi_k,[J_j,J_{j+\ell+1}(\tau)]\Phi_k\rangle.
\end{eqnarray}
Since the commutator $[J_j,\chi_{[j,j+\ell]}]$ is local, the contribution of the first term in the right-hand side  
is vanishing in the limit $T\rightarrow\infty$. By using Schwarz inequality, one can show that
the second and third terms are also vanishing in the limit because the current operator $J_j$ is local. 
Thus, the contribution of the first term in the right-hand side of (\ref{CorJchi}) is vanishing. 

\subsection{Estimating the second term in the right-hand side of (\ref{CorJchi})}

To begin with, we write 
$$
f_j(s):=\lim_{\ell\nearrow\infty}\lim_{L\nearrow\infty}i\sum_{k:E_k\le E_{\rm F}}
\langle \Phi_k,[J_j,\chi_{[j,j+\ell]}(s)\Phi_k\rangle.
$$
This limit exists thanks to (\ref{chijjell1s}). Then, one has 
$$
g_j(\eta,T)=\int_{-T}^0 ds\; e^{\eta s}f_j(s)
$$
{from} (\ref{gjetaTellN}) and (\ref{gjetaT}).

Let $\Delta\alpha$ be a real number whose absolute value $|\Delta\alpha|$ is small. 
Note that 
$$
g_j((1+\Delta\alpha)\eta,T)-g_j(\eta,T)
=\int_{-T}^0 ds\;(e^{\Delta\alpha\eta s}-1)e^{\eta s}f_j(s).
$$
Further, we introduce 
$$
\tilde{g}_j(\eta,T):=\int_{-T}^0ds\; \eta se^{\eta s}f_j(s),
$$
in order to estimate the second term in the right-hand side of (\ref{CorJchi}). 
This quantity is nothing but the contribution corresponding to the second term. From these two equations, one has 
\begin{equation}
\label{derivgj}
\frac{g_j((1+\Delta\alpha)\eta,T)-g_j(\eta,T)}{\Delta\alpha}-\tilde{g}_j(\eta,T)
=\int_{-T}^0 ds\; \sum_{n=2}^\infty \frac{1}{n!}(\Delta\alpha)^{n-1} (\eta s)^ne^{\eta s}f_j(s). 
\end{equation}
In order to estimate this right-hand side, we write 
$$
F_j(s):=-\int_s^0 d\tau\; e^{\eta\tau/2}f_j(\tau)
$$
for short. Clearly, $F_j(s)=-g_j(\eta/2,-s)$, and hence this is uniformly bounded with respect to $\eta$ and $s$ 
{from} the assumption (\ref{assumgj}) of Proposition~\ref{pro:gjCCcor}. 
By integration by parts, the right-hand side of (\ref{derivgj}) can be calculated as  
\begin{eqnarray*}
& &\sum_{n=2}^\infty \frac{1}{n!}\int_{-T}^0 ds\; (\Delta\alpha)^{n-1} (\eta s)^n e^{\eta s}f_j(s)\\
&=&\sum_{n=2}^\infty \frac{1}{n!}\left[(\Delta\alpha)^{n-1} (\eta s)^n e^{\eta s/2}F_j(s)\right]_{-T}^0\\
&-&\sum_{n=2}^\infty \frac{1}{n!}\int_{-T}^0 ds\; 
\left[(\Delta\alpha)^{n-1}n\eta (\eta s)^{n-1}+(\Delta\alpha)^{n-1}(\eta s)^{n}\cdot\frac{\eta}{2}\right]
e^{\eta s/2}F_j(s).
\end{eqnarray*}
Since the first sum in the right-hand side is vanishing in the limit $T\nearrow\infty$, one has 
\begin{eqnarray*}
& &\left|\int_{-\infty}^0 ds\; \sum_{n=2}^\infty \frac{1}{n!}(\Delta\alpha)^{n-1} (\eta s)^n 
e^{\eta s}f_j(s)\right|\\
&\le& \int_{-\infty}^0 ds\cdot \eta\; 
\left[\sum_{n=2}^\infty \frac{1}{(n-1)!}|\Delta\alpha|^{n-1}|\eta s|^{n-1}+\frac{1}{2}\sum_{n=2}^\infty \frac{1}{n!}
|\Delta\alpha|^{n-1}|\eta s|^{n}\right]
e^{\eta s/2}|F_j(s)|\\
&\le& |\Delta\alpha| \int_{-\infty}^0 ds\cdot \eta\;\left[|\eta s|e^{|\Delta\alpha \eta s|}
+\frac{1}{2}|\eta s|^2 e^{|\Delta\alpha \eta s|}\right]e^{\eta s/2}|F_j(s)|\\
&\le& \mathcal{C}_1|\Delta\alpha|\int_{-\infty}^0 ds'\;\left[|s'|+\frac{1}{2}|s'|^2\right]e^{|\Delta\alpha||s'|}e^{s'/2}
\le {\rm Const.}|\Delta\alpha|,
\end{eqnarray*}
where we have used the assumption (\ref{assumgj}) with $F_j(s)=-g_j(\eta/2,-s)$. 
Combining this with (\ref{derivgj}), we obtain 
$$
\left|\lim_{T\nearrow\infty}\left[\frac{g_j((1+\Delta\alpha)\eta,T)-g_j(\eta,T)}{\Delta\alpha}-\tilde{g}_j(\eta,T)\right]\right|
\le {\rm Const.}|\Delta\alpha|.
$$
We recall one of the two assumptions of Proposition~\ref{pro:gjCCcor} which requires  
the existence of $\lim_{\eta\searrow 0}\lim_{T\nearrow\infty}g_j(\eta,T)$. 
{From} this assumption, one has 
$$
\lim_{\eta\searrow 0}\lim_{T\nearrow\infty}g_j((1+\Delta\alpha)\eta,T)=\lim_{\eta\searrow 0}\lim_{T\nearrow\infty}g_j(\eta,T).
$$
Combining this with the above inequality, we have 
$$
\left|\lim_{\eta\searrow 0}\lim_{T\nearrow\infty}\tilde{g}_j(\eta,T)\right|
\le {\rm Const.}|\Delta\alpha|.
$$
Since this holds for any small $|\Delta\alpha|$, we obtain the desired result, 
$$
\lim_{\eta\searrow 0}\lim_{T\nearrow\infty}\tilde{g}_j(\eta,T)=0. 
$$

\subsection{Estimating the third term in the right-hand side of (\ref{CorJchi})}

Using the identity (\ref{derivchis}), the third term in the right-hand side of (\ref{CorJchi}) is written as 
\begin{eqnarray}
& &-\int_{-T}^0 ds\; s
e^{\eta s}\langle \Phi_k,[J_j,\frac{d}{ds}\chi_{[j,j+\ell]}(s)]\Phi_k\rangle \nonumber \\
&=&-\int_{-T}^0 ds\; s
e^{\eta s}\langle \Phi_k,[J_j,J_{j,}(s)]\Phi_k\rangle
+\int_{-T}^0 ds\; s
e^{\eta s}\langle \Phi_k,[J_j,J_{j+\ell+1}(s)]\Phi_k\rangle.
\end{eqnarray}
The first term in the right-hand side is nothing but the desired contribution of the current-current correlation. 
Using Lieb-Robinson bound \cite{LR,NS,HK,NOS}, one can show that the contribution of the second term is vanishing 
in the limit $\ell\rightarrow\infty$. 

\Section{Proof of Theorem~\ref{theorem:periodic}}
\label{Appnd:PoofTheoremPeriodic}

In this appendix, we prove that the conductance is quantized to the universal conductance 
for the periodic potential $v$ under the assumptions in Theorem~\ref{theorem:periodic}.   

{From} the assumptions, we can choose the length $L$ of the chain to be $L=pM$ with a large positive integer $M$, 
where $p$ is the period of the periodic potential, 
and we impose the periodic boundary condition for the unperturbed Hamiltonian $H_0$. 
Since the validity of the relation (\ref{gj=JCorCor}) in the present case 
is proved in Appendix~\ref{appendix:ProofconduRelationPeriodic}, 
it is enough to calculate the right-hand side of  (\ref{gj=JCorCor}). 

We write $\Phi_{m,k}$ for the eigenvectors of the unperturbed Hamiltonian $H_0$ with the energy eigenvalue $E_{m,k}$, 
where $m$ and $k$ are, respectively, the band index and the wavenumber.  
Then, one has 
\begin{equation}
\label{hatgjetaN}
\hat{g}_j(\eta,L)=\sum_{m,k:E_{m,k}\le E_{\rm F}}\; \sum_{m',k':E_{m',k'}>E_{\rm F}}
\frac{4\eta(E_{m',k'}-E_{m,k})}{[(E_{m',k'}-E_{m,k})^2+\eta^2]^2}|\langle\Phi_{m,k},J_j\Phi_{m',k'}\rangle|^2
\end{equation}
{from} the definition (\ref{gjCCcor}) of $\hat{g}_j(\eta,L)$. 

To begin with, we note that $|\langle\Phi_{m,k},J_j\Phi_{m',k'}\rangle|=\mathcal{O}(M^{-1})$ 
because the current operator $J_j$ is local. 
Let $\delta$ be a small positive number. Then, the contributions of the double sums in (\ref{hatgjetaN}) 
which satisfy $E_{m',k'}-E_{m,k}\ge \delta$ are vanishing in the double limit $\eta\searrow 0$ and $M\nearrow\infty$.   
Actually, one has 
$$
\frac{\eta|E_{m',k'}-E_{m,k}|}{[(E_{m',k'}-E_{m,k})^2+\eta^2]^2}\le\frac{\eta|E_{m',k'}-E_{m,k}|}{|E_{m',k'}-E_{m,k}|^4}
\le \frac{\eta}{\delta^3}.  
$$
Combining this with the above $|\langle\Phi_{m,k},J_j\Phi_{m',k'}\rangle|=\mathcal{O}(M^{-1})$, 
one can easily show that the corresponding contributions are vanishing. 
Thus, it is enough to treat the energies near the Fermi energy $E_{\rm F}$.   
Without loss of generality, it is sufficient to deal with the following four cases:  
\begin{description} 
\item{(i)} $k=k_{\rm F}-\kappa, \quad k'=k_{\rm F}+\kappa'$
\item{(ii)} $k=-k_{\rm F}+\kappa, \quad k'=-k_{\rm F}-\kappa'$
\item{(iii)} $k=k_{\rm F}-\kappa, \quad k'=-k_{\rm F}-\kappa'$
\item{(iv)} $k=-k_{\rm F}+\kappa, \quad k'=k_{\rm F}+\kappa'$, 
\end{description}
where $k_{\rm F}$ is the Fermi wavenumber, and $\kappa$ and $\kappa'$ are a positive small variable. 

Consider first Case~(i). The energies near the Fermi energy are given by  
$$
E_{m,k}\sim E_{\rm F}+\frac{d}{dk}E_{m,k}(k_{\rm F})\cdot (-\kappa)=E_{\rm F}-v_{\rm F}\kappa
$$
and
$$
E_{m,k'}\sim E_{\rm F}+\frac{d}{dk}E_{m,k}(k_{\rm F})\cdot \kappa'=E_{\rm F}+v_{\rm F}\kappa'.
$$
Therefore, the difference is 
$$
E_{m,k'}-E_{m,k}\sim v_{\rm F}(\kappa'+\kappa).
$$

Because of the translational invariance, the eigenvectors can be written as 
$$
\Phi_{m,k}(n)=\frac{1}{\sqrt{M}}e^{ikr}u_{m,k}(q)
$$
in Bloch form, where we have written $n=pr+q$ with $r=0,1,2,\ldots,M-1$ and $q=1,2,\ldots,p$. 
The matrix elements in the right-hand side of (\ref{hatgjetaN}) are estimated as 
\begin{eqnarray*}
\langle \Phi_{m,k},J_j\Phi_{m,k'}\rangle\sim 
\langle \Phi_{m,k_{\rm F}},J_j\Phi_{m,k_{\rm F}}\rangle&=&
\frac{1}{M}\sum_{r=0}^{M-1}\langle \Phi_{m,k_{\rm F}},J_{j+pr}\Phi_{m,k_{\rm F}}\rangle\\
&\sim&\frac{1}{M}\langle u_{m,k_{\rm F}},\tilde{J}(k_{\rm F})u_{m,k_{\rm F}}\rangle\\
&=&\frac{1}{M}\langle u_{m,k_{\rm F}},\frac{\partial\tilde{H}_0(k_{\rm F})}{\partial k}u_{m,k_{\rm F}}\rangle\\
&=&\frac{1}{M}\frac{d E_{m,k}}{d k}(k_{\rm F})=\frac{1}{M}v_{\rm F},
\end{eqnarray*}
where $\tilde{J}$ and $\tilde{H}_0$ are, respectively, the Fourier transform.   
Substituting these into the right-hand side of (\ref{hatgjetaN}), the corresponding contribution is written 
$$
g_+(\eta):=\frac{v_{\rm F}^2}{\pi^2}\int_0^\delta d\kappa \int_0^\delta d\kappa'\;
\frac{v_{\rm F}(\kappa+\kappa')\eta}{[v_{\rm F}^2(\kappa+\kappa')^2+\eta^2]^2}
$$
in the limit $M\nearrow\infty$. Changing the variables in the integrals as 
 $\tilde{\kappa}=v_{\rm F}\kappa/\eta$ and $\tilde{\kappa}'=v_{\rm F}\kappa'/\eta$, 
one has 
$$
g_+(0)=\frac{1}{\pi^2}\int_0^\infty d\tilde{\kappa} \int_0^\infty d\tilde{\kappa'}\;
\frac{\tilde{\kappa}+\tilde{\kappa}'}{[(\tilde{\kappa}+\tilde{\kappa}')^2+1]^2}
$$
in the limit $\eta\searrow 0$. Further, by setting $w=\tilde{\kappa}+\tilde{\kappa}'$, we obtain 
$$
g_+(0)=\frac{1}{\pi^2}\int_0^\infty dw\; \frac{w^2}{(w^2+1)^2}=\frac{1}{4\pi}. 
$$  
Although this is only the leading term, all of the higher corrections can be proved to be vanishing 
in the limit $\eta\searrow 0$ by relying on the fact that the cutoff $\delta$ can be chosen 
to be an arbitrary small positive number. 

Similarly, one can deal with Case (ii). The resulting contribution for the conductance is given by 
$$
g_-(0)=\frac{1}{4\pi}.
$$

In order to treat Cases (iii) and (iv), we note that 
$$
\langle\Phi_{m,k_{\rm F}},J_j\Phi_{m,-k_{\rm F}}\rangle=
(-i)\left[\Phi_{m,k_{\rm F}}^\ast(j-1)\Phi_{m,-k_{\rm F}}(j)-\Phi_{m,k_{\rm F}}^\ast(j)\Phi_{m,-k_{\rm F}}(j-1)\right], 
$$
which is obtained from the expression (\ref{defJj}) of the current operator $J_j$. 
Since the Hamiltonian $H_0$ is a real symmetric matrix, one has 
$\Phi_{m,k_{\rm F}}^\ast(n)=\Phi_{m,-k_{\rm F}}(n)$ for all the sites $n$.  
Substituting this into the above right-hand side yields the vanishing of the matrix element.  
Therefore, the corresponding contributions for the conductance is vanishing, too, in the same way. 

Adding all the contributions, we obtain the desired result, 
$$
\lim_{\eta\searrow 0}\lim_{M\nearrow \infty}\hat{g}_j(\eta,N)=g_+(0)+g_-(0)=\frac{1}{2\pi}.
$$

\Section{Proof of (\ref{gj=JCorCor}) in the case of the periodic potentials}
\label{appendix:ProofconduRelationPeriodic}

In this appendix, we prove that the equality (\ref{gj=JCorCor}) is valid in the case of the periodic potentials 
without relying on the assumptions of Proposition~\ref{pro:gjCCcor}. 
For this purpose, it is sufficient to show that the contribution of the second term in the right-hand side of 
(\ref{CorJchi}) is vanishing because the contribution of the first term in the right-hand side of (\ref{CorJchi}) 
can be proved to be vanishing without relying on the assumptions of Proposition~\ref{pro:gjCCcor}, 
as shown in Appendix~\ref{firsttermCorJchi}.  

To begin with, we note that 
\begin{eqnarray}
& &\int_{-T}^0 ds\; s e^{\eta s}e^{i(E_{m',k'}-E_{m,k})s} \nonumber \\
&=&\left[\frac{iT}{E_{m,k}-E_{m',k'}+i\eta}-\frac{1}{(E_{m,k}-E_{m',k'}+i\eta)^2}
\right]\exp[-\eta T-i(E_{m',k'}-E_{m,k})T] \nonumber \\
& &+\frac{1}{(E_{m,k}-E_{m',k'}+i\eta)^2}. 
\label{integralOsc}
\end{eqnarray}
Since the first term in the right-hand side is vanishing in the limit $T\nearrow\infty$, we will treat only the second 
term in the following. The corresponding contribution is written 
\begin{eqnarray*}
& &\sum_{m,k:E_{m,k}\le E_{\rm F}}\int_{-\infty}^0 ds\; \eta s
e^{\eta s}\langle \Phi_{m,k},[J_j,\chi_{[j,j+\ell]}(s)]\Phi_{m,k}\rangle \\
&=&\eta \sum_{m,k:E_{m,k}\le E_{\rm F}}\; \sum_{m',k':E_{m',k'}>E_{\rm F}}\Biggl\{
\langle\Phi_{m,k},J_j\Phi_{m',k'}\rangle\langle\Phi_{m',k'},\chi_{[j,j+\ell]}\Phi_{m,k} \rangle 
\frac{1}{[E_{m,k}-E_{m',k'}+i\eta]^2}\\
& &\qquad\qquad\qquad -\langle\Phi_{m,k},\chi_{[j,j+\ell]}\Phi_{m',k'}\rangle\langle\Phi_{m',k'},J_j\Phi_{m,k} \rangle 
\frac{1}{[E_{m,k}-E_{m',k'}-i\eta]^2}\Biggr\}. 
\end{eqnarray*}
The matrix element of $\chi_{[j,j+\ell]}$ in the right-hand side can be calculated as 
$$
\langle\Phi_{m',k'},\chi_{[j,j+\ell]}\Phi_{m,k} \rangle  
=\frac{1}{M}e^{ij(k-k')}\frac{e^{i(\hat{\ell}+1)(k-k')}-1}{e^{i(k-k')}-1}\langle u_{m',k'},u_{m,k}\rangle, 
$$
where we have taken $\ell=p\hat{\ell}$ with a large positive integer $\hat{\ell}$. 

In the same way as in Appendix~\ref{Appnd:PoofTheoremPeriodic}, it is sufficient to treat 
Cases~(i) and (ii). In the following, we consider only Case~(i) because both of Cases can be treated in the same way. 
In the former case, the above matrix element can be written as 
$$
\langle\Phi_{m,k'},\chi_{[j,j+\ell]}\Phi_{m,k} \rangle
\sim \frac{i}{M}\frac{e^{-i(\hat{\ell}+1)(\kappa+\kappa')}-1}{\kappa+\kappa'}.
$$
Therefore, the corresponding contribution is 
$$
\tilde{g}_j(\eta,\hat{\ell})=\frac{v_{\rm F}}{4\pi^2}\int_0^\delta d\kappa \int_0^\delta d\kappa'
\frac{\eta}{\kappa+\kappa'}\left\{
\frac{e^{-i(\hat{\ell}+1)(\kappa+\kappa')}-1}{[v_{\rm F}(\kappa+\kappa')-i\eta]^2}
+\frac{e^{i(\hat{\ell}+1)(\kappa+\kappa')}-1}{[v_{\rm F}(\kappa+\kappa')+i\eta]^2}\right\}.
$$
In the double limit $\hat{\ell}\nearrow\infty$ and $\eta\searrow 0$, we have 
$$
\lim_{\eta\searrow 0}\lim_{\hat{\ell}\nearrow \infty}
\tilde{g}_j(\eta,\hat{\ell})=-\frac{1}{4\pi^2}\int_0^\infty dw\left\{\frac{1}{(w-i)^2}+\frac{1}{(w+i)^2}\right\}=0,
$$
where we have used Riemann-Lebesgue argument for $\hat{\ell}\nearrow\infty$.  

Finally, we remark the following: 
We can expect that the uniform bound (\ref{assumgj}) in Proposition~\ref{pro:gjCCcor} holds  
in the case of periodic potentials because the oscillatory integrals which appear in the contributions 
from the first term in the right-hand side of (\ref{CorJchi}) and the first term in the right-hand side of (\ref{integralOsc}) 
cancel the factor $T$ which appears in their contributions for a large $T$.  
In fact, one can prove this statement under certain assumptions about the spectrum
of the Hamiltonian $H_0$ and its Fourier transform.


\end{document}